\documentclass[aps,nofootinbib,amsfonts,superscriptaddress,showpacs,showkeys]{revtex4-1}
\usepackage{amsfonts,amssymb,amscd,amsmath}
\usepackage{graphicx}
\usepackage{subfigure}
\usepackage{appendix}
\graphicspath{{figures/}}
\allowdisplaybreaks
\newcommand{\rr}{\mbox{\boldmath $r$}}
\newcommand{\rp}{\mbox{\boldmath $r^{\prime}$}}
\newcommand{\ba}{\mbox{\boldmath $a$}}
\newcommand{\bE}{\mbox{\boldmath $E$}}

\newcommand{\be}{\mbox{\boldmath $e$}}
\newcommand{\bj}{\mbox{\boldmath $j$}}
\newcommand{\brho}{\mbox{\boldmath $\rho$}}
\newcommand{\bsigma}{\mbox{\boldmath $\sigma$}}

\newcommand{\pv}{\mbox{\boldmath $p$}}
\newcommand{\lv}{\mbox{\boldmath $l$}}
\newcommand{\kv}{\mbox{\boldmath $k$}}
\newcommand{\bv}{\mbox{\boldmath $b$}}
\newcommand{\sv}{\mbox{\boldmath $s$}}
\newcommand{\Bv}{\mbox{\boldmath $B$}}
\newcommand{\Deltav}{\mbox{\boldmath $\Delta$}}

\DeclareMathOperator{\Tr}{Tr}

\newcommand{\ket}[1]{{#1} \rangle}
\newcommand{\bra}[1]{\langle {#1} }

\newcommand{\dd}{\, \mathrm{d}}
\newcommand{\rb}{\mbox{\boldmath $b$}}
\begin{document}
\title{Associated  jet + electroweak gauge boson production in hadronic collisions  \\ at forward rapidities  in the  color -  dipole $S$ - matrix  framework}

\author{Yan B. {\sc Bandeira}}
\email{yan.bandeira@ufpel.edu.br}
\affiliation{Institute of Physics and Mathematics, Federal University of Pelotas, \\
  Postal Code 354,  96010-900, Pelotas, RS, Brazil}
\affiliation{Institute of Nuclear Physics PAN, PL-31-342 Cracow, Poland}

\author{Victor P. {\sc Gon\c{c}alves}}
\email{barros@ufpel.edu.br}
%\affiliation{Institut f\"ur Theoretische Physik, Westf\"alische Wilhelms-Universit\"at M\"unster,
%Wilhelm-Klemm-Stra\ss e 9, D-48149 M\"unster, Germany}
\affiliation{Institute of Physics and Mathematics, Federal University of Pelotas, \\
  Postal Code 354,  96010-900, Pelotas, RS, Brazil}
%\affiliation{Institute of Modern Physics, Chinese Academy of Sciences,  Lanzhou 730000, China}

\author{Wolfgang {\sc Sch\"afer}}
\email{Wolfgang.Schafer@ifj.edu.pl}
\affiliation{Institute of Nuclear Physics PAN, PL-31-342 Cracow, Poland}

\begin{abstract}
The cross-section for the associated  production of a jet with an electroweak gauge boson ($G = W^{\pm}, Z^0,  \gamma$)   at forward rapidities in $pp$ and $pA$ collisions is derived within the  color -  dipole $S$ - matrix framework. We present the full expressions for the differential cross-section of the $q p \rightarrow G q X$ process in the transverse momentum space, considering the longitudinal and transverse polarizations of the gauge boson.  We demonstrate that the final formulae can be expressed in terms of the  unintegrated gluon distribution and reproduce previous results for the associated jet  + $\gamma$ and jet + $Z^0$ production, derived using other frameworks. { Moreover, we derive the back - to - back correlation limit of the spectra and show that it can be expressed in terms of the unpolarized and linearly polarized transverse momentum gluon distributions.} Our results improve the description of the inclusive jet plus color neutral particle production at forward rapidities, not far from the proton fragmentation region, in $pp$ or $pA$ collisions,  and are the main ingredient to study the impact of nonlinear QCD effects in two - particle correlations.
\end{abstract}
\maketitle

%%%%%%%%%%%%%
\section{Introduction}

The production of electroweak gauge bosons  ($G = W^{\pm}, Z^0,  \gamma$)  in hadronic collisions is one of the more promising observables to investigate the high energy regime of  quantum chromodynamics (QCD) {(For a review see, e.g. Ref. \cite{Tricoli:2020uxr})}. 
The absence of final-state interactions makes this process a clean probe of the  wave functions of the incident particles, and this characteristic has motivated a large number of phenomenological studies over the last decades \cite{Aurenche:1988vi, Vogelsang:1995bg, Yang:2022qgk,Gehrmann-DeRidder:2019avi,BrennerMariotto:2007yf,BrennerMariotto:2008st,Arleo:2011gc,dEnterria:2012kvo,Helenius:2014qla,Klasen:2017dsy,Goharipour:2018sip,Kopeliovich:2007yva,Kopeliovich:2009yw,SampaiodosSantos:2020lte,Gelis:2002ki,Jalilian-Marian:2012wwi,Ducloue:2017kkq,Goncalves:2020tvh,Lima:2023dqw,Kopeliovich:2000fb,Raufeisen:2002zp,Kopeliovich:2001hf,Betemps:2004xr,Betemps:2003je,Golec-Biernat:2010dup,Ducati:2013cga,Schafer:2016qmk,Ducloue:2017zfd,Gelis:2002fw,Baier:2004tj,Stasto:2012ru,Kang:2012vm,Basso:2016ulb,Basso:2015pba,Marquet:2019ltn,Benic:2016uku,Benic:2017znu,Benic:2018hvb,Golec-Biernat:2020cah,Benic:2022ixp,Taels:2023czt}.
In particular, this process has been largely studied considering the production of a gauge boson at forward rapidities, where the partons from the projectile scatter off a dense gluonic system in the target and  the cross-section is expected to be  sensitive to nonlinear effects in the QCD dynamics  \cite{Kopeliovich:2007yva,Kopeliovich:2009yw,SampaiodosSantos:2020lte,Gelis:2002ki,Jalilian-Marian:2012wwi,Ducloue:2017kkq,Goncalves:2020tvh,Lima:2023dqw,Kopeliovich:2000fb,Raufeisen:2002zp,Kopeliovich:2001hf,Betemps:2004xr,Betemps:2003je,Golec-Biernat:2010dup,Ducati:2013cga,Schafer:2016qmk,Ducloue:2017zfd,Gelis:2002fw,Baier:2004tj,Stasto:2012ru,Kang:2012vm,Basso:2016ulb,Basso:2015pba,Marquet:2019ltn,Benic:2016uku,Benic:2017znu,Benic:2018hvb,Golec-Biernat:2020cah}.
In a recent paper \cite{Bandeira:2024zjl},  we have derived using the $S$ - matrix framework \cite{Nikolaev:1994de,Nikolaev:1995ty,Nikolaev:2003zf,Nikolaev:2004cu,Nikolaev:2005dd,Nikolaev:2005zj,Nikolaev:2005ay,Nikolaev:2005qs}, the general formulae for the inclusive electroweak gauge boson production at forward rapidities and demonstrated that it reduces to those used in Refs. \cite{Kopeliovich:2007yva,Kopeliovich:2009yw,SampaiodosSantos:2020lte,Gelis:2002ki,Jalilian-Marian:2012wwi,Ducloue:2017kkq,Goncalves:2020tvh,Lima:2023dqw,Kopeliovich:2000fb,Raufeisen:2002zp,Kopeliovich:2001hf,Betemps:2004xr,Betemps:2003je,Golec-Biernat:2010dup,Ducati:2013cga,Schafer:2016qmk,Ducloue:2017zfd,Gelis:2002fw,Baier:2004tj,Stasto:2012ru,Kang:2012vm,Basso:2016ulb,Basso:2015pba,Marquet:2019ltn} to estimate the real photon and $Z^0$ production in the appropriate limits and representations.  Moreover, we have derived, for the first time, the cross-section for the $W^{\pm}$ production in the hybrid factorization formalism. Such results are an important improvement in the description of the inclusive electroweak gauge boson production and are useful to investigate the impact of the nonlinear effects in the transverse momentum and rapidity distributions of the gauge boson probed in kinematical range of the LHCb detector. 

A more detailed information about the QCD dynamics is expected to be provided through the study of processes where two particles are tagged in the final state (See, e.g., Ref. \cite{Morreale:2021pnn} for a recent review). Such expectation is directly associated with the presence in the dense target of a characteristic momentum scale -- the saturation scale $Q_s$ --, which implies that the incident partons acquire a momentum imbalance due to its multiple scattering with the target \cite{hdqcd}. As a consequence, the nonlinear effects are expected to generate a depletion of the back - to - back peak predicted by the collinear formalism \cite{Marquet:2007vb,Albacete:2010pg,Stasto:2011ru,Stasto:2018rci}. Such an expectation have been confirmed by the analysis performed in Refs. \cite{Dominguez:2011wm,Stasto:2012ru,Jalilian-Marian:2012wwi,Basso:2015pba,Ducloue:2017kkq,Goncalves:2020tvh,Taels:2023czt}, which have estimated the azimuthal angle correlation for  the associated hadron + $\gamma$ and hadron +$Z^0$ production considering different frameworks and distinct approximations in the calculation of the $q p \rightarrow qGX$ cross-section.       
Our goal, in this paper, is to extend the formalism presented in Ref. \cite{Bandeira:2024zjl} for the associated jet plus electroweak gauge boson production  at forward rapidities in hadronic collisions and present the general formulae for the 
$q p \rightarrow qGX$ cross-section considering  the longitudinal and transverse polarizations of the gauge boson. As we will demonstrate below, our results reproduce the expressions used in Refs. \cite{Dominguez:2011wm,Stasto:2012ru,Jalilian-Marian:2012wwi,Basso:2015pba,Ducloue:2017kkq,Goncalves:2020tvh,Benic:2022ixp,Taels:2023czt} in the appropriated limits. Moreover, we will derive, for the first time, the cross-section for the associated jet + $W^{\pm}$ production. Our results will be presented in the transverse momentum representation, which implies that the cross-section will be expressed in terms of the unintegrated gluon distribution of the target, {which is related with dipole gluon distribution and that resums both the initial and final state interactions \cite{Dominguez:2011wm}}.  Therefore, our results can be directly used to investigate the impact of the description of QCD dynamics at high energies on the associated jet plus electroweak gauge boson production at LHC. { Moreover, we investigate the back - to - back correlation limit, in which the mean dijet momentum is much larger than the recoil momentum, and demonstrate that the spectrum can be expressed in terms of the unpolarized and linearly
polarized transverse momentum gluon distributions.}

This paper is organized as follows. In the next section, we will review the main results obtained in Ref. \cite{Bandeira:2024zjl} and derive the general formulae for the differential $q p \rightarrow qGX$ cross-section in the transverse momentum representation.  The vector and axial contributions will be estimated, and the expressions for the longitudinal and transverse polarizations will be explicitly presented. In section \ref{sec:particular}, we will present the explicit results for the distinct gauge bosons and discuss some particular cases of our general expression. In particular, we will demonstrate that it reduces, in some specific limits, to the cross-sections used in the literature to calculate the associated jet + $\gamma$ and jet + $Z^0$ production. {The back - to - back correlation limit is investigated in section \ref{sec:correlation}.}  Finally, in section \ref{sec:conc}, our main conclusions are summarized and prospects are discussed. 

%Two appendixes are also included, where the double logarithmic limit of our results is discussed  and some useful integrals are shown.

%The forward hadron production in hadron-hadron collisions is a typical example of a dilute-dense process, which is an ideal system to study the small-x components of the target wave function.

%============================

\begin{figure}[t]
{\includegraphics[width=0.6\textwidth]{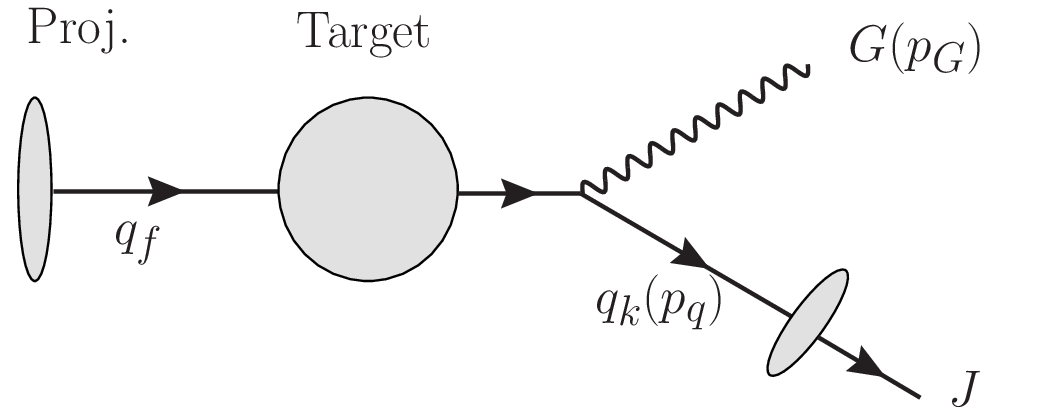}}  
%{\includegraphics[width=0.3\textwidth]{Gaugeboson_jet.eps}}                                                                                  
\caption{Typical diagram contributing for the associated $G$ + jet production in hadronic collisions, where the  gauge boson is irradiated by a quark  of flavor $f$  after  the interaction with the target color field (denoted by a shaded circle). The quark $q_k$ generates the jet $J$. For the $W^{\pm}$ radiation $q_k \neq q_f$.}
\label{fig:diagram}
\end{figure}   
   
%===========================

\section{Associated jet + electroweak gauge boson production}
\label{sec:formalism_dijet}
The treatment of the two - particle production at forward rapidities in hadronic collisions at high energies and the derivation of the corresponding cross-sections is still a challenge, especially when both particles in the final state are generated by partons, since in this case it depends in a non-trivial way on  quadrupole correlators of fundamental Wilson lines or  four partons $S$ matrix, depending on the approach considered estimating this quantity \cite{Nikolaev:2005qs,Dominguez:2011wm}. The calculation of these objects in the general case  is a hard task. In contrast, if one of the particles in the final state is an electroweak gauge boson, which does not interact strongly, these quantities can be expressed in terms of dipole correlators or two partons $S$ matrix. As  explicitly demonstrated e.g. in our previous paper  
 \cite{Bandeira:2024zjl}, the cross-sections  for the isolated electroweak gauge boson production {in $pp$ collisions} can be expressed in terms of the unintegrated gluon distribution or, equivalently, of the dipole - proton cross-section, which can be constrained using experimental data for $ep$ collisions. Such characteristic motivates a more detailed investigation of the associated jet + electroweak gauge boson production. 
 
The hadronic cross-section for the associated jet + $G$ production, represented in Fig. \ref{fig:diagram}, can be  schematically expressed as follows
 %(For a recent review, see e.g. Ref. \cite{vanHameren:2023oiq}):
\begin{eqnarray}
d\sigma(h_A h_B \rightarrow G \, J \, X) \propto  q_{f/A} \otimes d\sigma(q_f h_B \rightarrow q_k G X) \,\,,
%\otimes D_{H_1/b} \otimes D_{H_2/c} \,\,,
\end{eqnarray}
i.e., as a convolution of the standard
quark distributions for the dilute projectile, $q_{f/A}$, and the parton - target cross-section, $d\sigma(q_f h_B \rightarrow q_k G X)$, which includes the nonlinear QCD effects, with the quark $q_k$ generating the jet $J$.
 In the  color - dipole $S$ - matrix framework for hadronic collisions, proposed in Refs. \cite{Nikolaev:1994de,Nikolaev:1995ty} and generalized in a series of publications \cite{Nikolaev:2003zf,Nikolaev:2004cu,Nikolaev:2005dd,Nikolaev:2005zj,Nikolaev:2005ay,Nikolaev:2005qs}, the $q_f h_B \rightarrow q_k G X$ process  can be viewed  as an excitation of the perturbative $|G q_k \rangle$ Fock state of the physical projectile $|q_f\rangle$ by multiple gluon exchanges with the target \cite{Nikolaev:1994de,Nikolaev:1995ty}. At high energies, the quark $q_f$ can be assumed to propagate along a straight line with a fixed impact parameter and the perturbative transition $q_f \rightarrow G q_k$ can be described in terms of the Fock state expansion for the physical state $|q_f\rangle_{phys}$. Such an expansion allows deriving  the action of the $S$ matrix on $|q_f\rangle_{phys}$, which will be given by \cite{Nikolaev:2003zf}
\begin{eqnarray}
S|q_f\rangle_{phys} & = & S_{q_f}(\rb)|q_f\rangle_{0} + S_G(\rb_G)S_{q_k}(\rb_{q})\Psi(z,\rr)|G{q_k}\rangle \nonumber \\
& = & S_{q_f}(\rb)|q_f\rangle_{phys} + [S_G(\rb_G)S_{q_k}(\rb_{q})-S_{q_f}(\rb)]\Psi(z,\rr)|G{q_k}\rangle \,\,,
\label{Eq:Smatrix}
\end{eqnarray}
where $|...\rangle_0$ refers to bare quarks and $\Psi(z,\rr)$ is the probability amplitude to find the $Gq_k$ system with separation $\rr$ in the two-dimensional impact parameter space and $z$ the fraction of the longitudinal momentum of incoming quark $q_f$ carried by the gauge boson $G$. The corresponding scattering amplitude can be expressed as follows 
\cite{Nikolaev:2004cu}
\begin{eqnarray}
{\cal{A}} = \int d^2\rb_G \, d^2\rb_{q} \exp[-i(\pv_G \cdot \rb_G + \pv_{q} \cdot \rb_{q})]\, [S_G(\rb_G)S_{q_k}(\rb_{q})-S_{q_f}(\rb)]\Psi(z,\rr)\,\,, 
\label{eq:amplitude}
\end{eqnarray}
where $\rb$ is  the impact parameter of the incoming quark $q_f$, $\rb_G = \rb + (1-z)\rr$ for the gauge boson and $\rb_{q} = \rb - z\rr$ for the quark $q_k$. Moreover, $\pv_G$ and $\pv_{q}$ are the transverse momenta of the gauge boson and quark in the final state, respectively. Therefore, the   differential cross-section for the associated jet plus gauge boson production in the color - dipole $S$ - matrix framework is given by \cite{Bandeira:2024zjl}
\begin{eqnarray}
\label{Eq:quark+G}
& \, & \frac{d\sigma^f_{T,L} (q_f h_B \rightarrow G(p_G) q_k(p_q))}{dz d^2\pv_G d^2\pv_q}  =  \frac{1}{(2\pi)^4}\, 
\int d^2\bv_G d^2\bv_q d^2\bv_G^{\prime} d^2\bv_q^{\prime} \exp[-i\pv_G \cdot (\bv_G - \bv_G^{\prime}) - i\pv_q \cdot (\bv_q - \bv_q^{\prime}) ]\, \nonumber \\
& \times & \Psi_{T,L}(z,\bv_G - \bv_q) \Psi^{*}_{T,L}(z,\bv_G^{\prime} - \bv_q^{\prime}) 
 \left\{S^{(2)}_{q\bar{q}}(\bv_q^{\prime},\bv_q) + S^{(2)}_{q\bar{q}}(\bv^{\prime},\bv) - S^{(2)}_{q\bar{q}}(\bv,\bv_q^{\prime}) - S^{(2)}_{q\bar{q}}(\bv^{\prime},\bv_q) \right\} \,, 
\end{eqnarray}
%%%%
where $S_{q\bar{q}}^{(2)}(\rb^\prime,\rb)$ represents the $S$ matrix for the interaction of the $q_f\bar{q}_k$ state with the target, with $\bar{q}_k$ propagating at the impact parameter $\rb^\prime$. Conservation of orbital angular momentum leads to the relations $\bv = z \bv_G + (1-z) \bv_q$ and  $\bv' = z \bv_G' + (1-z) \bv'_q$
between the impact parameters of incoming and outgoing partons in amplitude and complex conjugate amplitude, respectively. The other impact parameters can be expressed as follows,
\begin{eqnarray}
\bv_G &=& \bv + (1-z) \rr \, , \, \bv_q = \bv - z \rr \nonumber \\
\bv'_G &=& \bv' + (1-z) \rr' \, , \, \bv'_q = \bv' - z \rr' \, .
\end{eqnarray}
with $\rr = \bv_G - \bv_q$ and $\rr' = \bv'_G - \bv'_q$. As a consequence, the differential cross-section can be written as
%%%%
\begin{eqnarray}
\label{eq:Master_formula_reduced}
\frac{d\sigma^f_{T,L} (q_f \rightarrow G(p_G) q_k(p_q))}{dz d^2\pv_G d^2\pv_q} & = & \frac{1}{(2\pi)^4}\, 
\int  d^2\rr d^2\rr'  \exp[- i ((1-z)\pv_G- z \pv_q) \cdot (\rr - \rr^{\prime}) ] \Psi_{T,L}(z,\rr) \Psi^{*}_{T,L}(z,\rr') \nonumber \\
&\times&\int d^2\bv d^2\bv'\exp[-i (\pv_G + \pv_q) \cdot (\bv - \bv^{\prime})] \nonumber \\
&\times&
 \left\{S^{(2)}_{q\bar{q}}(\bv'-z\rr',\bv-z\rr) + S^{(2)}_{q\bar{q}}(\bv^{\prime},\bv) - 
S^{(2)}_{q\bar{q}}(\bv'-z\rr',\bv) - S^{(2)}_{q\bar{q}}(\bv^{\prime},\bv-z\rr) \right\}  \,\,,
\end{eqnarray}
%%%%
which indicates that  the conjugate variable to  $\rr - \rr'$ is the light-cone relative momentum $\pv = (1-z) \pv_G - z \pv_q$. Moreover, $L$ ($T$) refers to the longitudinal (transverse) polarization of the gauge boson.
%%%%

In order to derive  the  differential cross-section in terms of the dipole - proton cross-section, it is useful to define the variables  $\sv = \bv - \bv'$ and $\Bv = (\bv + \bv')/2$, as well as the  transverse momentum decorrelation $\Deltav = \pv_G + \pv_q$. Performing the change of variables and the integration over $\Bv$, { %summing over the gauge boson polarizations and 
averaging over the quark polarizations and summing over polarizations of transverse gauge bosons}, results in 
%that
%%%%
\begin{eqnarray}
%\label{Eq:Master_dijet1}
\frac{d\sigma^f_{T,L} (q_f \rightarrow G(p_G) q_k(p_q))}{dz d^2\pv d^2\Deltav} & = & \frac{1}{2 (2\pi)^4}\, 
\int  d^2\rr d^2\rr'  \exp[- i \pv \cdot (\rr - \rr^{\prime}) ] \overline{\sum_\text{pol.}} \Psi_{T,L}(z,\rr) \Psi^{*}_{T,L}(z,\rr') \nonumber \\
&\times&\int d^2\sv \exp[- i \Deltav \cdot \sv] %\nonumber \\
%&\times&
 \Big\{
 \sigma_{q\bar{q}}(\sv+z\rr') + \sigma_{q\bar{q}}(\sv-z\rr) 
- \sigma_{q\bar{q}}(\sv -z(\rr-\rr') ) - \sigma_{q\bar{q}}(\sv)
\Big\}  \,\,, 
\end{eqnarray}
%%%%%
where we have used that:
%%%
\begin{eqnarray}
\sigma_{q \bar q}(\rr) = 2 \int d^2\Bv \, \Big[ 1 - S^{(2)}_{q \bar q} \Big(\Bv + {\rr \over 2}, \Bv - {\rr \over 2} \Big)\Big] \, .
\label{Eq:dip}
\end{eqnarray}
The light front wave functions (LFWFs) for the $q \to G q'$ transitions have been calculated in light front gauge in Ref. 
\cite{Bandeira:2024zjl}, and they imply
\begin{eqnarray} \label{Psi2}
%%&  & \overline{\sum}_\text{pol.} \Psi_{T,L}(z,\rr,m_f,m_k) \Psi^{*}_{T,L}(\,\rp,m_f,m_k)  =  \rho^V_{T,L}(z,\rr,m_f,m_k) + \rho^A_{T,L}(z,\rr,m_f,m_k) \,.
&  & \overline{\sum_\text{pol.}} \, \Psi_{T,L}(z,\rr) \Psi^{*}_{T,L}(z,\rp)  =  \rho^V_{T,L}(z,\rr,\rr') + \rho^A_{T,L}(z,\rr,\rr') \,,
\end{eqnarray}
with the vector($V$) and axial ($A$) contributions being given by
%$\rho^V_{T,L}$ and $\rho^A_{T,L}$ for the vector ($V$) and axial ($A$) contributions were derived in Ref.  \cite{Bandeira:2024zjl},   and are given by
\begin{subequations} \label{eq:rho-defs}
\begin{align}
\rho^T_V(z,\rr,\rr') &= \frac{({\cal C}^G_f)^2(g^{G}_{V,f})^2}{2\pi^2}  \Big\{ \frac{1 + (1-z)^2}{z} \, \frac{\rr \cdot \rr'}{rr'} \epsilon^2 {\rm K}_1(\epsilon r) {\rm K}_1(\epsilon r') + z\Big((m_k - m_f) + z m_f\Big)^2 {\rm K}_0(\epsilon r) {\rm K}_0(\epsilon r') \Big\} \label{eq:rho-def-TV} \\  
\rho^T_A(z,\rr,\rr') &= \frac{({\cal C}^G_f)^2(g^{G}_{A,f})^2}{2\pi^2}  \Big\{ \frac{1 + (1-z)^2}{z} \, \frac{\rr \cdot \rr'}{rr'} \epsilon^2 {\rm K}_1(\epsilon r) {\rm K}_1(\epsilon r') + z\Big((m_k + m_f) - z m_f\Big)^2 {\rm K}_0(\epsilon r) {\rm K}_0(\epsilon r') \Big\} \label{eq:rho-def-TA} \\
\begin{split}\label{eq:rho-def-LV}
\rho^L_V(z,\rr,\rr') &= \frac{({\cal C}^G_f)^2(g^{G}_{V,f})^2}{4\pi^2} 
\Big\{ \frac{(z^2 m_f(m_k - m_f) - z (m_k^2 - m_f^2) - 2 (1-z) M_G^2)^2}{z M_G^2} \, {\rm K}_0(\epsilon r) {\rm K}_0(\epsilon r')  \\
& + \frac{z (m_k - m_f)^2}{M_G^2}  \, \frac{\rr \cdot \rr'}{rr'} \, \epsilon^2 {\rm K}_1(\epsilon r) {\rm K}_1(\epsilon r') \Big\}     
\end{split}\\
\begin{split}\label{eq:rho-def-TA}
\rho^L_A(z,\rr,\rr') &= \frac{({\cal C}^G_f)^2(g^{G}_{A,f})^2}{4\pi^2}  
\Big\{ \frac{(z^2 m_f(m_k + m_f) + z (m_k^2 - m_f^2) + 2 (1-z) M_G^2)^2}{z M_G^2} \, {\rm K}_0(\epsilon r) {\rm K}_0(\epsilon r')  \\
& + \frac{z (m_k + m_f)^2}{M_G^2} \, \frac{\rr \cdot \rr'}{rr'} \, \epsilon^2 {\rm K}_1(\epsilon r) {\rm K}_1(\epsilon r') \Big\} \,\,,  
\end{split}
\end{align}
\end{subequations}
where $g^G_{V,f}$ and $g^G_{A,f}$ are the vector and axial couplings for the distinct gauge bosons and quark flavors, $C_f^G$ the associated coefficients (For details see Ref. \cite{Bandeira:2024zjl}), $\epsilon^2 =  (1-z) M_G^2 + z (m_f^2 - m_k^2) + z^2 m_k^2$,  $m_f$ and $m_k$ are the masses of the incoming and outgoing quarks, respectively, $M_G$ is the mass of the gauge boson and the functions ${\rm K}_0$ and ${\rm K}_1$ are the modified Bessel functions.
%In the next section, we will use these results in Eq. (\ref{Eq:isolated}) to derive the expressions for the spectrum associated with the isolated gauge boson production. Such expressions reduce to  those derived in Refs.  \cite{kst99,Pasechnik:2012ac}  for $G = \gamma^*$ and $Z^*$ using different frameworks.

Finally, using the relation between the dipole - proton cross-section and the unintegrated gluon distribution $f(x,\kv)$ given by
\begin{eqnarray}
\sigma_{q \bar q}(\rr)=\int\dd^{2}\kv f(x,\kv)\left(1-e^{i\kv\cdot\rr}\right)\,\,,
\label{Eq:unint}
\end{eqnarray}
the spectrum for the jet + $G$ production  in the transverse momentum representation will be expressed by
%\begin{equation}
%\label{Eq:Master_mom_dijet}
%\begin{split}
%\frac{d\sigma^f_{T,L} (q_f \rightarrow G(p_G) q_k(p_q))}{dz d^2\pv d^2\Deltav} & =&  \frac{1}{2 (2\pi)^4}\,
%\int  d^2\rr d^2\rr'  \, e^{i \pv \cdot (\rr - \rr^{\prime})}\, \big[\rho^{T,L}_{V}(z,\rr,\rr') + \rho^{T,L}_{A}(z,\rr,\rr')\big]  \int d^2\sv \,  e^{i \Deltav \cdot \sv} \\
%&\times
% \Big\{
%e^{i\kv\cdot\sv} + e^{i\kv\cdot(\sv-z(\rr-\rp))}
%- e^{i\kv\cdot(\sv+z\rp)}-e^{i\kv\cdot(\sv-z\rr)}
%\Big\}  \,  f(x,\kv)  \,. 
%\end{split}
%\end{equation}

\begin{eqnarray}
\label{Eq:Master_mom_dijet}
%\begin{split}
\frac{d\sigma^f_{T,L} (q_f \rightarrow G(p_G) q_k(p_q))}{dz d^2\pv d^2\Deltav} & =&  \frac{1}{2 (2\pi)^4}\,
\int d^2 \kv \, \int  d^2\rr d^2\rr'  \, e^{-i \pv \cdot (\rr - \rr^{\prime})}\, \big[\rho^{T,L}_{V}(z,\rr,\rr') + \rho^{T,L}_{A}(z,\rr,\rr')\big] \nonumber \\
&& \int d^2\sv \,  e^{-i \Deltav \cdot \sv} 
%\\
%&\times&
 \Big\{
e^{i\kv\cdot\sv} + e^{i\kv\cdot(\sv-z(\rr-\rp))}
- e^{i\kv\cdot(\sv+z\rp)}-e^{i\kv\cdot(\sv-z\rr)}
\Big\}  \,  f(x,\kv) 
%\nonumber \\
%&=&  f(x, \Deltav) \frac{1}{2 (2\pi)^2}\,
%\int  d^2\rr d^2\rr'  \, e^{i \pv \cdot (\rr - \rr^{\prime})}\, \big[\rho^{T,L}_{V}(z,\rr,\rr') + \rho^{T,L}_{A}(z,\rr,\rr')\big] \nonumber \\
%&\times&  \Big\{
%1 + e^{- i z \Deltav\cdot(\rr-\rp)}
%- e^{i z \Deltav\cdot \rp}-e^{- i z\Deltav\cdot\rr}
%\Big\} 
%\,. 
%\end{split}
\end{eqnarray}
%%%%
Equation (\ref{Eq:Master_mom_dijet}) is the main ingredient to estimate the associated jet plus electroweak gauge boson production in the small-$x$ limit. {One has that the behavior of the spectrum is strongly dependent on the  unintegrated gluon distribution $f(x,\kv)$, which is implicitly defined through the dipole cross-section 
according to Eq. (\ref{Eq:unint}). Such quantity is directly related to what in the literature is known as the   ``dipole gluon distribution'' \cite{Dominguez:2011wm}, which sums up initial- and final-state interactions relative to the hard process. }

%%%%
\section{Particular cases}
\label{sec:particular}

In this section, we will give the explicit expressions for the differential cross-sections for the associated jet + electroweak gauge boson production, considering separately the distinct gauge bosons. In particular, we will derive, for the first time, the differential spectrum for the associated jet + $W^{\pm}$ production. Moreover, we will demonstrate that Eq. (\ref{Eq:Master_mom_dijet}) reproduces, in the appropriated limits,  the formulae previously used in the literature to estimate the associated jet + $\gamma$ and jet + $Z^0$ production at forward rapidities in hadronic collisions.

\subsection{Associated jet + $W^{\pm}$ production}
For this particular case, the initial and final quarks have different flavors and both the axial and vector contributions contribute to the cross-section. The master equation, Eq. (\ref{Eq:Master_mom_dijet}), { implies that the spectrum for a transverse polarization will be expressed by}
%\begin{equation}
%\label{Eq:Master_mom_jetW}
%\begin{split}
%\frac{d\sigma^f_{T,L} (q_f \rightarrow W^{\pm} q_k)}{dz d^2\pv d^2\Deltav} & =  \frac{1}{2 (2\pi)^4}\,
%\int  d^2\rr d^2\rr'  \, e^{i \kv \cdot (\rr - \rr^{\prime})}\,  %\big[\rho^{T,L}_{V}(z,\rr,\rr') + \rho^{T,L}_{A}(z,\rr,\rr')\big]
 %\int d^2\kv f(x,\kv)
 %\\
%&\times
%\int d^2\sv \,  e^{i \Deltav \cdot \sv} 
% \Big\{
%e^{i\kv\cdot\sv} + e^{i\kv\cdot(\sv-z(\rr-\rp))}
%- e^{i\kv\cdot(\sv+z\rp)}-e^{i\kv\cdot(\sv-z\rr)}
%\Big\}  \,\,.     
%\end{split}
%\end{equation}
% For the transverse polarization, one has that  
\begin{eqnarray}
\frac{d\sigma^f_{T}(q_f \rightarrow W^{\pm} q_k)}{d z d^2\pv d^2 \Deltav } & = & \frac{1}{2(2\pi)^4}\,\int d^2\rr \dd^2\rp e^{-i\pv \cdot (\rr - \rp)}
\rho^T_V(z,\rr,\rr') \int d^2\kv f(x,\kv)\,
\nonumber \\ 
&\times&
\int d^2\sv \,  e^{-i \Deltav \cdot \sv}
\Big\{
e^{i\kv\cdot\sv} + e^{i\kv\cdot(\sv-z(\rr-\rp))}
- e^{i\kv\cdot(\sv+z\rp)}-e^{i\kv\cdot(\sv-z\rr)}
\Big\} \nonumber \\
& + & \frac{1}{2(2\pi)^4}\,\int d^2\rr \dd^2\rp e^{-i\pv \cdot (\rr - \rp)}
\rho^T_A(z,\rr,\rr') \int d^2\kv f(x,\kv)\, 
\nonumber \\ 
&\times&
\int d^2\sv \,  e^{-i \Deltav \cdot \sv}
 \Big\{
e^{i\kv\cdot\sv} + e^{i\kv\cdot(\sv-z(\rr-\rp))}
- e^{i\kv\cdot(\sv+z\rp)}-e^{i\kv\cdot(\sv-z\rr)}
\Big\}.
\end{eqnarray}
Using the expression for the squared LFWF, Eq. (\ref{eq:rho-defs}), one gets
% {\bf{- Verified and corrected in Oct 11 (YB). great, thanks! WS}}
\begin{eqnarray}\label{dsigTV-prods}
    \frac{d\sigma^f_{T}(q_f \rightarrow W^{\pm} q_k)}{dz d^2\pv d^2 \Deltav}\Bigg|_{V} & = & \frac{1}{2(2\pi)^4}\,
    \frac{\big(C^{W^+}_{f}\big)^{2}\big(g^{W^+}_{V,f}\big)^{2}}{2\pi^{2}}
    \int d^2\kv f(x,\kv)
    \Bigg\{
    \int d^2\rr \dd^2\rp e^{-i\pv \cdot (\rr - \rp)}
    \int d^2\sv \,  e^{-i \Deltav \cdot \sv} \nonumber \\ 
    &\times& \frac{1 + (1-z)^2}{z} \, \frac{\rr \cdot \rr'}{rr'} \epsilon^2 {\rm K}_1(\epsilon r) {\rm K}_1(\epsilon r') 
    \Big(
    e^{i\kv\cdot\sv} + e^{i\kv\cdot(\sv-z(\rr-\rp))}
    - e^{i\kv\cdot(\sv+z\rp)}-e^{i\kv\cdot(\sv-z\rr)}
    \Big)
    \nonumber \\ 
    &+& 
    \int d^2\rr \dd^2\rp e^{-i\pv \cdot (\rr - \rp)}
    \int d^2\sv \,  e^{-i \Deltav \cdot \sv} \,
    z\Big((m_k - m_f) + z m_f\Big)^2 {\rm K}_0(\epsilon r) {\rm K}_0(\epsilon r') \nonumber \\
    &\times&
    \Big(
    e^{i\kv\cdot\sv} + e^{i\kv\cdot(\sv-z(\rr-\rp))}
    - e^{i\kv\cdot(\sv+z\rp)}-e^{i\kv\cdot(\sv-z\rr)}
    \Big)   \Bigg\}\, ,
\end{eqnarray}
where $C_{f}^{W^{\pm}}=\frac{\sqrt{\alpha_{em}}}{2\sqrt{2}\sin\theta_{W}}V^{\pm}$ and $g_{A,V,f}^{W^{\pm}}=1$, for $V^{+}=V_{f_{u},f_{d}}$ and $V^{-}=V_{f_{d},f_{u}}$, which corresponds to the CKM matrix elements for  up-type quarks ($f_u =u,c,t$) and  down-type quarks ($f_d=d,s,b$).  Using the results presented in Appendix C of Ref.\cite{Bandeira:2024zjl}
%\ref{ap: integrals mom space}, 
one gets:
%{\bf sth. strange about factors of $2\pi$ below, also we should pull out the delta-function. (The $\pi$'s factors were wrong because I did not take into account the $2(2\pi)^4$ inside the curly brackets, now it is correct. Sorry about this! - YB)}
\begin{equation}
    \begin{split}
    \frac{d\sigma^f_{T}(q_f \rightarrow W^{\pm} q_k)}{dz d^2\pv d^2 \Deltav}\Bigg|_{V} & = \frac{\big(C^{W^+}_{f}\big)^{2}\big(g^{W^+}_{V,f}\big)^{2}}{(2\pi)^6}
    \int\dd^{2}\kv f(x,\kv) \delta^{(2)}(\Deltav-\kv)  \\
    & \times
    \left\{ \frac{1+(1-z)^{2}}{z}\epsilon^{2}\left[-\frac{2(2\pi)^{4}}{\epsilon^{2}}\frac{\pv\cdot(\pv-z\kv)}{(p^{2}+\epsilon^{2})[(\pv-z\kv)^{2}+\epsilon^{2}]}\right.\right. \\ 
    &+
    \left.\frac{(2\pi)^{4}}{\epsilon^{2}}\frac{(\pv-z\kv)^{2}}{[(\pv-z\kv)^{2}+\epsilon^{2}]^{2}}+\frac{(2\pi)^{4}}{\epsilon^{2}}\frac{p^{2}}{(p^{2}+\epsilon^{2})^{2}}\right] \\
    &+ 
    z\left[(m_{k}-m_{f})+zm_{f}\right]^{2}\left(-\frac{2(2\pi)^{4}}{(p^{2}+\epsilon^{2})[(\pv-z\kv)^{2}+\epsilon^{2}]}\right. \\
    &+
    \left.\left.\frac{(2\pi)^{4}}{[(\pv-z\kv)^{2}+\epsilon^{2}]^{2}}+\frac{(2\pi)^{4}}{(p^{2}+\epsilon^{2})}\right)\right\} 
    \end{split}
%\begin{split}
%    \frac{d\sigma^f_{T}q_f \rightarrow W^{\pm} q_k)}{dz d^2\pv d^2 \Deltav}\Bigg|_{V} & = \frac{\big(C^{W^+}_{f}\big)^{2}\big(g^{W^+}_{V,f}\big)^{2}}{(2\pi)^6}
%    \int\dd^{2}\kv f(x,\kv) \\
%    & \times
%    \left\{ \frac{1+(1-z)^{2}}{z}\epsilon^{2}\left[-\frac{2(2\pi)^{4}}{\epsilon^{2}}\frac{\pv\cdot(\pv-z\kv)\delta^{(2)}(\Deltav-(-\kv))}{(p^{2}+\epsilon^{2})[(\pv-z\kv)^{2}+\epsilon^{2}]}\right.\right. \\ 
%    &+
%    \left.\frac{(2\pi)^{4}}{\epsilon^{2}}\frac{(\pv-z\kv)^{2}}{[(\pv-z\kv)^{2}+\epsilon^{2}]^{2}}\delta^{(2)}(\Deltav-(-\kv))+\frac{(2\pi)^{4}}{\epsilon^{2}}\frac{p^{2}}{(p^{2}+\epsilon^{2})^{2}}\delta^{(2)}(\Deltav-(-\kv))\right] \\
%   &+ 
%  z\left[(m_{k}-m_{f})+zm_{f}\right]^{2}\left(-\frac{2(2\pi)^{4}}{(p^{2}+\epsilon^{2})[(\pv-z\kv)^{2}+\epsilon^{2}]}\delta^{(2)}(\Deltav-(-\kv))\right. \\
%    &+
%    \left.\left.\frac{(2\pi)^{4}}{[(\pv-z\kv)^{2}+\epsilon^{2}]^{2}}\delta^{(2)}(\Deltav-(-\kv))+\frac{(2\pi)^{4}}{(p^{2}+\epsilon^{2})}\delta^{(2)}(\Deltav-(-\kv))\right)\right\} 
%   \end{split}
\end{equation}
which can be rewritten as 
\begin{equation}
\begin{split}
    \frac{d\sigma^f_{T}(q_f \rightarrow W^{\pm} q_k)}{dz d^2\pv d^2 \Deltav}\Bigg|_{V}
    &=
\frac{\big(C^{W^+}_{f}\big)^{2}\big(g^{W^+}_{V,f}\big)^{2}}{2\pi^{2}}
    f(x,\Deltav)\Bigg\{ \frac{1+(1-z)^{2}}{z}\mathcal{E}_{2}(\pv,z\Deltav,\epsilon)  \\
    &  + z\left[(m_{k}-m_{f})+zm_{f}\right]^{2}\mathcal{E}_{1}(\pv,z\Deltav,\epsilon)\Bigg\} 
\end{split}
\end{equation}
where we have defined the  auxiliary functions: 
%{\bf in the previous paper they were called ${\cal E}$}.
\begin{eqnarray}
%{\cal{T}}_2(\pv,\Deltav,\epsilon,z) &\equiv& 2 
{\cal{E}}_1(\pv,\Deltav,\epsilon) 
 \equiv \frac{1}{2}
\left[ \frac{1}{p^2 + \epsilon^2} - \frac{1}{(\pv- \Deltav)^2 + \epsilon^2}  \right]^2.
\end{eqnarray}
\begin{eqnarray}
%{\cal{T}}_1(\pv,\Deltav,\epsilon,z) &\equiv& 2 
{\cal{E}}_2(\pv,z\Deltav,\epsilon)
 \equiv \frac{1}{2}
\left[ \frac{\pv}{p^2 + \epsilon^2} - \frac{\pv - z\Deltav}{(\pv- z\Deltav)^2 + \epsilon^2}  \right]^2,
\end{eqnarray}

Similarly, the axial contribution for a transverse polarization, can be expressed by:
\begin{eqnarray}
    \frac{d\sigma^f_{T}(q_f \rightarrow W^{\pm} q_k)}{dz d^2\pv d^2 \Deltav}\Bigg|_{A}
    &=&
    \frac{\big(C^{W^+}_{f}\big)^{2}\big(g^{W^+}_{A,f}\big)^{2}}{2\pi^{2}}
    f(x,\Deltav)\left\{ \frac{1+(1-z)^{2}}{z}\mathcal{E}_{2}(\pv,z\Deltav,\epsilon) + z\left[(m_{k}+m_{f})-zm_{f}\right]^{2}\mathcal{E}_{1}(\pv,z\Deltav,\epsilon)\right\} \nonumber \\
\end{eqnarray}

Following similar steps, one can derive that the vector and axial contributions for the longitudinal cross-section will be given by 
%{\bf factors of $2 pi$? (It's correct this way - YB)}
\begin{eqnarray}
   \frac{d\sigma^f_{L}(q_f \rightarrow W^{\pm} q_k)}{dz d^2\pv d^2 \Deltav}\Bigg|_{V}
    &=& 
\frac{\big(C^{W^+}_{f}\big)^{2}\big(g^{W^+}_{V,f}\big)^{2}}{(2\pi)^{2}}
    f(x,\Deltav)\left\{ \frac{z(m_{k}-m_{f})^{2}}{M_{W}^{2}}\mathcal{E}_{2}(\pv,z\Deltav,\epsilon)\right. \nonumber\\
    &+&
    \left.\frac{\left[z^{2}m_{f}(m_{k}-m_{f})-z(m_{k}^{2}-m_{f}^{2})-2(1-z)M_{W}^{2}\right]^{2}}{zM_{W}^{2}}\mathcal{E}_{1}(\pv,z\Deltav,\epsilon)\right\} .
\end{eqnarray}
and
\begin{eqnarray}
    \frac{d\sigma^f_{L}(q_f \rightarrow W^{\pm} q_k)}{dz d^2\pv d^2 \Deltav}\Bigg|_{A}
    &=& \frac{\big(C^{W^+}_{f}\big)^{2}\big(g^{W^+}_{A,f}\big)^{2}}{(2\pi)^{2}}
    f(x,\Deltav)\left\{ \frac{z(m_{k}+m_{f})^{2}}{M_{W}^{2}}\mathcal{E}_{2}(\pv,z\Deltav,\epsilon)\right. \nonumber\\
    &+&
    \left.\frac{\left[z^{2}m_{f}(m_{k}+m_{f})+z(m_{k}^{2}-m_{f}^{2})+2(1-z)M_{W}^{2}\right]^{2}}{zM_{W}^{2}}\mathcal{E}_{1}(\pv,z\Deltav,\epsilon)\right\} 
\end{eqnarray}

The above expressions are the main input to estimate the associated jet + $W^{\pm}$ production at forward rapidities in hadronic collisions. These results indicate that the differential cross-section is strongly sensitive to the description of the unintegrated gluon distribution, and that this quantity determines the momentum decorrelation between the jet and the gauge boson.

\subsection{Associated Jet +  photon production}
\label{subsec:gamma_jet}

For the photon production, the  axial contribution vanishes, since $g_{A,f}^{\gamma^{*}}=0$. Moreover, $g_{V,f}^{\gamma^{*}}=1$, $C_{V,f}^{\gamma^{*}}=\sqrt{\alpha_{em}}e_{f}$ and $m_{k}=m_{f}$ once the emitting quark does not change its flavor. Thus, the differential cross-section for this final state is given by
\begin{eqnarray}
\left.\frac{d \sigma}{d z d^{2}\pv d^{2}\Deltav}\right|_{qp\rightarrow\gamma^{*}X} &=&	\left.\frac{\dd\sigma_{T}}{\dd z\dd^{2}\pv\dd^{2}\Deltav}\right|_{V}+\left.\frac{\dd\sigma_{L}}{\dd z\dd^{2}\pv\dd^{2}\Deltav}\right|_{V}, \nonumber \\
\left.\frac{d \sigma}{d z d^{2}\pv d^{2}\Deltav}\right|_{qp\rightarrow\gamma^{*}X} &=&	\frac{\alpha_{em}e_{f}^{2}}{2\pi^{2}}f(x,\Deltav)\left\{ \frac{1+(1-z)^{2}}{z}\mathcal{E}_{2}(\pv,z\Deltav,\epsilon)+z^{3}m_{f}^{2}\mathcal{E}_{1}(\pv,z\Deltav,\epsilon)\right\} \nonumber \\
&+&
\frac{\alpha_{em}e_{f}^{2}}{4\pi^{2}}f(x,\Deltav)\left\{ \frac{4(1-z)^{2}M_{\gamma}^{2}}{z}\mathcal{E}_{1}(\pv,z\Deltav,\epsilon)\right\}  \nonumber \\
\left.\frac{d \sigma}{d z d^{2}\pv d^{2}\Deltav}\right|_{qp\rightarrow\gamma^{*}X} &=&		\frac{\alpha_{em}e_{f}^{2}}{2\pi^{2}}f(x,\Deltav)\left\{ \frac{1+(1-z)^{2}}{z}\mathcal{E}_{2}(\pv,z \Deltav,\epsilon)\right.
+\left.\left[z^{3}m_{f}^{2}+\frac{2(1-z)^{2}M_{\gamma}^{2}}{z}\right]\mathcal{E}_{1}(\pv,z\Deltav,\epsilon)\right\},
\end{eqnarray}
 where $\epsilon^{2}=(1-z)M_{\gamma}^{2}+z^{2}m_{f}^{2}$ in the general case of a virtual photon. 
 In the massless quark case, one has
  \begin{eqnarray}
    \left.\frac{d \sigma}{d z d^{2}\pv d^{2}\Deltav}\right|_{qp\rightarrow\gamma^{*}X}^{m_{f}=0}
    &=&\frac{\alpha_{em}e_{f}^{2}}{2\pi^{2}}f(x,\Deltav)\left\{ \frac{1+(1-z)^{2}}{z}\mathcal{E}_{2}(\pv,z \Deltav,\epsilon)+\frac{2(1-z)^{2}M_{\gamma}^{2}}{z}\mathcal{E}_{1}(\pv,z\Deltav,\epsilon)\right\},    
  \end{eqnarray}
which can be expressed as follows 
%({\bf after some manipulation of $\mathcal E_2$})
\begin{eqnarray}
\left. z \, \frac{d \sigma}{d z d^{2}\pv  d^{2}\Deltav}\right|_{qp\rightarrow\gamma^{*}X}^{m_{f}=0}
&=&\frac{\alpha_{em}e_{f}^{2}}{(2\pi)^{2}}f(x,\Deltav)\left\{ \left[ 1+(1-z)^{2} \right]\frac{z^{2}\Deltav^{2}}{\left((\pv-z\Deltav)^{2}+(1-z) M_\gamma^2\right)\left(\pv^{2}+(1-z)M_\gamma^{2}\right)}\right. \nonumber \\
&-& \left.z^{2}(1-z)M_{\gamma}^{2}\left[\frac{1}{\left((\pv-z\Deltav)^{2}+(1-z)M_\gamma^{2}\right)}-\frac{1}{\left(\pv^{2}+(1-z) M_\gamma^{2}\right)}\right]^{2}\right\} .
\label{eq:virtual}
\end{eqnarray}
%{\bf - Verified and corrected in Oct 11 (YB) Very good, thanks Yan! Was this a typo just here? The eqns. 10a)-10c) taken from our previous paper are ok? I am checking some hard matrix elements in the collinear limit comparing with what is in the literature. It was just one $z$ missing, that now is corrected. The equations 10a) and 10c) are ok. (YB - 14 Oct) {\bf Thanks! Btw, sorry for temporarily messing up the style of latex, for the moment it is easier for me to work this way...}}
Finally, for a real photon ($M_{\gamma}^{2}=0$) and massless quarks,  we obtain
\begin{eqnarray}
z\, \left.\frac{d \sigma}{d z d^{2}\pv d^{2}\Deltav}\right|_{qp\rightarrow \gamma X}
&=&\frac{\alpha_{em}e_{f}^{2}}{(2\pi)^{2}}f(x,\Deltav) \left[1+(1-z)^{2}\right]\frac{z^{2}\Deltav^{2}}{(\pv-z\Deltav)^{2} \pv^{2}} .
\label{eq:real}
\end{eqnarray}
In this case one needs to take care of the collinear pole in the final state $q \to q \gamma$ splitting at $\pv = z\Deltav$, for example by imposing a photon isolation condition.
The Eqs. (\ref{eq:virtual}) and (\ref{eq:real}) agree with those used in Refs.\cite{Dominguez:2011wm,Stasto:2012ru,Basso:2015pba,Taels:2023czt} to estimate the jet + photon production in hadronic collisions.

\subsection{Associated  Jet + $Z^0$ production}

For the $Z^0$ production, the axial and vector contributions must be included, and the total differential cross-section will be given by
\begin{eqnarray}
    \left.\frac{d \sigma}{d z d^{2}\pv d^{2}\Deltav}\right|_{qp\rightarrow Z^0 X} &=&	\left.\frac{\dd\sigma_{T}}{\dd z\dd^{2}\pv\dd^{2}\Deltav}\right|_{V}
    +\left.\frac{\dd\sigma_{T}}{\dd z\dd^{2}\pv\dd^{2}\Deltav}\right|_{A}
    +\left.\frac{\dd\sigma_{L}}{\dd z\dd^{2}\pv\dd^{2}\Deltav}\right|_{V}
    +\left.\frac{\dd\sigma_{L}}{\dd z\dd^{2}\pv\dd^{2}\Deltav}\right|_{A}. 
\end{eqnarray}
Moreover, one has that $ {\cal C}^Z_f=\frac{\sqrt{\alpha_{em}}}{\sin 2\theta_W}$ and $g^Z_{V,A}$ depends on the quark flavor. For up-type flavors ($f_u=u,c,t$) $g_{V,f_u}^Z=\frac12-\frac43\sin^2\theta_W$ and $g_{A,f_u}^Z=\frac12$, while for down-type flavors ($f_d=d,s,b$) $g_{V,f_d}^Z=-\frac12+\frac23\sin^2\theta_W $ and $g_{A,f_d}^Z=-\frac12$. Also, $m_k=m_f$, as this kind of process does not imply a change of quark flavor. Then, the parton level cross-sections for each contribution will be given by:
\begin{eqnarray}
    \frac{d\sigma^f_{T}}{dz d^2\pv d^2 \Deltav}\Bigg|_{V}^{Z^0}
    &=&
    \frac{(C_{f}^{Z})^{2}(g^Z_{V,f})^{2}}{2\pi^{2}} f(x,\Deltav)\left\{ \frac{1+(1-z)^{2}}{z}\mathcal{E}_{2}(\pv,z\Deltav,\epsilon) + z^3m_f^2\mathcal{E}_{1}(\pv,z\Deltav,\epsilon)\right\} \\
    \frac{d\sigma^f_{T}}{dz d^2\pv d^2 \Deltav}\Bigg|_{A}^{Z^0}
    &=&
    \frac{(C_{f}^{Z})^{2}(g^Z_{A,f})^{2}}{2\pi^{2}} f(x,\Deltav)\left\{ \frac{1+(1-z)^{2}}{z}\mathcal{E}_{2}(\pv,z\Deltav,\epsilon) +z(2-z)m^2_f\mathcal{E}_{1}(\pv,z\Deltav,\epsilon)\right\} \\
    \frac{d\sigma^f_{L}}{dz d^2\pv d^2 \Deltav}\Bigg|_{V}^{Z^0}
    &=& \frac{(C_{f}^{Z})^{2}(g_{V,f}^{Z})^{2}}{(2\pi)^{2}}f(x,\Deltav)\left\{ 
    4\frac{(1-z)^2}{z}M^2_Z\mathcal{E}_{1}(\pv,z\Deltav,\epsilon)\right\} \\
    \frac{d\sigma^f_{L}}{dz d^2\pv d^2 \Deltav}\Bigg|_{A}^{Z^0}
    &=& \frac{(C_{f}^{Z})^{2}(g_{A,f}^{Z^0})^{2}}{(2\pi)^{2}}f(x,\Deltav)\left\{ 4\frac{zm^2_f}{M_{Z}^{2}}\mathcal{E}_{2}(\pv,z\Deltav,\epsilon)
    +
    \frac{\left[4z^{2}m^2_f-2(1-z)M_{Z}^{2}\right]^{2}}{zM_{Z}^{2}}\mathcal{E}_{1}(\pv,z\Deltav,\epsilon)\right\} , 
\end{eqnarray}
where $\epsilon^{2}=(1-z)M_{Z}^{2}+z^{2}m_{f}^{2}$. In the massless quark case, the total cross-section can be expressed as follows, 
%{\bf need to replace $\epsilon^2$ below}
\begin{eqnarray}
    \left.\frac{d\sigma}{dzd^{2}\pv d^{2}\Deltav}\right|_{qp\rightarrow Z^{0}X}^{m_{f}=0} &=&\frac{(C_{f}^{Z})^{2}}{(2\pi)^{2}}\left[(g_{V,f}^{Z})^{2}+(g_{A,f}^{Z})^{2}\right]f(x,\boldsymbol{\Delta})\left\{ \frac{1+(1-z)^{2}}{z}\left[\frac{\boldsymbol{p}-z\boldsymbol{\Delta}}{[(\boldsymbol{p}-z\boldsymbol{\Delta})^{2}+\bar{\epsilon}^{2}]}-\frac{\boldsymbol{p}}{(p^{2}+\bar{\epsilon}^{2})}\right]^{2}\right. \nonumber   \\
    &&\left.+2\frac{(1-z)^{2}}{z}M_{Z}^{2}\left[\frac{1}{[(\boldsymbol{p}-z\boldsymbol{\Delta})+\bar{\epsilon}^{2}]}-\frac{1}{(p^{2}+\bar{\epsilon}^{2})}\right]^{2}\right\} \,\,,
\end{eqnarray}
{ with $\bar{\epsilon}^{2}=(1-z)M_{Z}^{2}$.} Such an expression was used in Refs. \cite{Basso:2015pba,Basso:2016ulb} to estimate the associated jet + $Z^0$ production at forward rapidities in hadronic collisions.

\section{The back - to - back correlation limit and the TMD factorization}
\label{sec:correlation}

Over the last years, the multiparticle production in electron - hadron and hadron - hadron collisions has been a topic of intense study, 
strongly motivated by the fact such a process is expected to be  sensitive to the physics of gluon saturation, via azimuthal correlations among the produced particles  (For a review  see, e.g., Ref. \cite{Albacete:2014fwa}). In particular, the inclusive production of a pair of jets has been largely studied using the Color Glass Condensate (CGC) formalism { (See, e.g., Refs. \cite{Marquet:2007vb,Albacete:2010pg,Stasto:2011ru,Dominguez:2011wm,Dumitru:2015gaa,Stasto:2018rci,Mantysaari:2019hkq,Caucal:2021ent,Caucal:2022ulg})}, resulting in the derivation of the general expressions for the cross - sections associated with different final states as well as in the demonstration that they exhibit the  transverse-momentum dependent (TMD) factorisation~\footnote{For a review about TMD factorization see, e.g., Ref. \cite{Boussarie:2023izj}.} in the correlation limit \cite{Dominguez:2011wm}. In this limit, the relative momentum is much larger than the transverse momentum decorrelation, i.e.  $ |\pv| \gg |\Deltav|$, which implies that the two measured jets are relatively hard and nearly back-to-back in the transverse plane.
Moreover, in this case, the cross - sections can be expressed in terms of the unpolarized and linearly polarized TMD gluon distributions
\cite{Metz:2011wb,Dominguez:2011br}, with the linearly polarized distribution contributing for azimuthal asymmetries. Our goal here is to investigate the correlation limit of the spectrum for  the associated jet + electroweak gauge boson production, derived in the previous sections using the color - dipole $S$ - matrix framework, and establish the connection of our results with the TMD factorization.

Let us now demonstrate how to obtain the TMD representation in the back-to-back correlation limit starting from  Eq.(\ref{eq:amplitude}). Expanding the $S$ - matrix one obtain that:
%%%%
\begin{eqnarray}
{\cal{A}} &=& \int d^2\rb \, d^2\rr \exp[-i\Deltav \cdot \rb -i \rr \cdot \pv]\, [S_{q}(\rb - z\rr )-S_{q}(\rb)]\Psi(z,\rr)\,\,,  \nonumber \\
&\approx& -z \int d^2\rb \, d^2\rr \exp[-i\Deltav \cdot \rb -i \rr \cdot \pv]\, (\rr \cdot \nabla_\perp  S_{q} \, (\rb) ) \Psi(z,\rr)\, ,
\end{eqnarray}
%%%%
so that the associated cross-section becomes
%%%%
\begin{eqnarray}
\frac{d\sigma^f_{T,L} (q_f \rightarrow G(p_G) q_k(p_q))}{dz d^2\pv d^2\Deltav} & = & \frac{z^2}{(2\pi)^4 \, N_c}\, 
\int  d^2\rr d^2\rr'  \exp[- i \pv \cdot (\rr - \rr^{\prime}) ] \overline{\sum_\text{pol.}} \Psi_{T,L}(z,\rr) r_i \Psi^{*}_{T,L}(z,\rr') r'_j \nonumber \\
&\times&\int d^2\rb d^2\rb' \exp[- i \Deltav \cdot (\rb - \rb')] %\nonumber \\
%&\time
\bra{N} | \Tr[\partial_\perp^i S_q(\rb) \partial_\perp^j S^\dagger (\rb') ] | \ket{N}  \, \nonumber \\
&\equiv& H_{ij}(z,\pv) \Phi_{ij}(x,\Deltav) \, .
\label{eq:correlation_limit}
\end{eqnarray}
%%%%
Here
%%%%
\begin{eqnarray}
H_{ij}(z,\pv) &=& 
\frac{z^2}{ (2\pi)^4 \, N_c}\, 
\int  d^2\rr d^2\rr'  \exp[- i \pv \cdot (\rr - \rr^{\prime}) ] \overline{\sum_\text{pol.}} \Psi_{T,L}(z,\rr) r_i \Psi^{*}_{T,L}(z,\rr') r'_j \, ,
\nonumber \\
&=& \frac{z^2}{ (2\pi)^4 \, N_c}\, 
\overline{\sum_\text{pol.}} \partial^i_{p\perp} \, \Psi_{T,L}(z,\pv) \, \partial_{p \perp}^j \Psi_{T,L}^*(z,\pv) \, ,
\end{eqnarray}
%%%
plays the role of the hard matrix element, and
%Defining the quantities
%%
%\begin{eqnarray}
%H_{ij}(z,\pv) &=& %\frac{z^2}{2 (2\pi)^4 \, N_c}\, 
%\int  d^2\rr d^2\rr'  \exp[- i \pv \cdot (\rr - \rr^{\prime}) ] \overline{\sum_\text{pol.}} \Psi_{T,L}(z,\rr) r_i \Psi^{*}_{T,L}(z,\rr') r'_j \, ,
%\nonumber \\
%&=& \frac{z^2}{2 (2\pi)^4 \, N_c}\, 
%\partial^i_{p\perp} \Psi(z,\pv) \, \partial_{p \perp}^j \Psi^*(z,\pv) \, ,
%\end{eqnarray}
%and
%%%%
\begin{eqnarray}
\Phi_{ij}(x,\Deltav) =  \int d^2 \bv d^2 \bv' \, 
\exp[- i \Deltav \cdot (\bv - \bv')] 
{\bra{N} | \Tr[\partial_\perp^i S_q(\rb) \partial_\perp^j S^\dagger (\rb') ] | \ket{N} } \, .
\label{eq:Phi_ij}
\end{eqnarray}
%%%%
is related to the target gluon TMD.
%which are related to the gauge boson emission and quark - target interaction, respectively, one has that
%the cross - section can be factorized as follows
%\begin{eqnarray}
%\frac{d\sigma^f_{T,L} (q_f \rightarrow G(p_G) q_k(p_q))}{dz d^2\pv d^2\Deltav} & = & 
% H_{ij}(z,\pv) \, \Phi_{ij}(x, \Deltav) \, . 
%\label{eq:correlation_limit2}
%\end{eqnarray}
Let us now consider the decomposition of $ \Phi_{ij}(x,\Deltav)$ into two orthogonal tensor structures, given by
%%%%
\begin{eqnarray}
    \Phi_{ij}(x,\Deltav) = \frac{1}{2} \delta^\perp_{ij} {\cal F}(x,\Deltav) + \frac{1}{2} \Big( 2 \frac{\Delta_i \Delta_j}{\Deltav^2} - \delta^\perp_{ij}  \Big) {\cal H}(x,\Deltav) \, .
\label{eq:tensor_decomp}
\end{eqnarray}
%%%
Using integration by parts in Eq. (\ref{eq:Phi_ij}), we %can easily show, that 
obtain 
%%%
\begin{eqnarray}
    \Phi_{ij}(x,\Deltav) = \Deltav_i \Deltav_j  \int d^2 \bv d^2 \bv' \, 
\exp[- i \Deltav \cdot (\bv - \bv')] 
{\bra{N} | \Tr[S_q(\rb) S^\dagger (\rb') ] | \ket{N} } \, .
\end{eqnarray}
%%%
We can now multiply this tensor by the appropriate tensors to project out ${\cal F},{\cal H}$, namely:
%%%%
\begin{eqnarray}
{\cal F}(x,\Deltav) = \delta^\perp_{ij} \, \Phi_{ij}(x,\Deltav), \quad {\cal H}(x,\Deltav) = \Big( 2 \frac{\Delta_i \Delta_j}{\Deltav^2} - \delta^\perp_{ij}  \Big) \, \Phi_{ij}(x,\Deltav) \, . 
\end{eqnarray}
%%%
This results in 
%%%
\begin{eqnarray}
{\cal{F}}(x,\Deltav) = {\cal H}(x,\Deltav) = \Deltav^2  \int d^2 \bv d^2 \bv' \, 
\exp[- i \Deltav \cdot (\bv - \bv')] 
{\bra{N} | \Tr[S_q(\rb) S^\dagger (\rb') ] | \ket{N} } \, .
\label{eq:F_and_H}
\end{eqnarray}
%%%%
Evidently, the correlator $\Phi_{ij}$ has only one independent component, which  plays the role of the TMD gluon distribution in the factorization formula Eq.(\ref{eq:correlation_limit}) for the back-to-back correlation limit.  
Furthermore, the tensor decomposition of Eq.(\ref{eq:tensor_decomp}) reminds the decomposition of the gluon TMD into an unpolarized  and linearly polarized distribution, { performed e.g.  in Refs. \cite{Metz:2011wb,Dominguez:2011wm}.
Following reference \cite{Dominguez:2011wm},} one can be more precise with the relation of $\Phi_{ij}$ to gluon TMDs. To this end, we introduce the Wilson line
%%%%
\begin{eqnarray}
U(\xi_f, \xi_i; \bv)  = P \,  \exp\Big( i g_s \int_{\xi_i}^{\xi_f} dx^+ \, A^{a -}(x^+,0,\bv) t^a \Big)  \, ,
\end{eqnarray}
%%%%
and observe that 
%%%
\begin{eqnarray}
 S_q(\bv) = U(\infty, -\infty; \bv) \, .     
\end{eqnarray}
%%%%
Then, one would write the derivative $\partial_\perp^i S_q(\bv)$ as
%%%%
\begin{eqnarray}
   \partial_\perp^i S_q(\bv) = i g_s \int_{- \infty}^{\infty} dx^+  U(\infty,x^+,\bv) \, \partial_\perp^i A^{-a}(x^+,0,\bv)t^a \, U(x^+,-\infty;\bv)  
\end{eqnarray}
%%%%
Inserting this expression into Eq.(\ref{eq:Phi_ij}), and identifying $\partial^i_\perp A^{-a}$ with the dominant component of the gluon field strength tensor
one would obtain that the function ${\cal F}(x,\Deltav)$ is related to the so-called dipole TMD where the Wilson lines combine into the staple-like form. 
The explicit relation to the dipole TMD $x G_{\rm dip}(x,\Deltav)$ commonly used in the literature (see e.g. Eq.(8.8)) of Ref. \cite{Boussarie:2023izj}) is
%%%%
\begin{eqnarray}
 {\cal F}(x, \Deltav) = 8 \pi^4 \alpha_s \, x G_{\rm dip}(x,\Deltav) \, ,
\end{eqnarray}
%%%%
In order to avoid ambiguities regarding factors of $\pi$, let us mention that these distributions are understood to be differential in $d^2 \Deltav$ so that the mass dimension of $\cal{F}$ is ${\rm GeV}^{-2}$.

We now remind the reader of the general form of the gauge boson production cross section in TMD factorization, which reads (omitting the convolution over initial state quark distributions):

\begin{eqnarray}
 \frac{d\sigma^f_{T,L} (q_f \rightarrow G(p_G) q_k(p_q))}{dz d^2\pv d^2\Deltav} = H_{qg \to G^*q}\, x f_1(x,\Deltav) + \cos(2 \phi) \, H^{cos(2 \phi)}_{qg \to G^* q}
\, x h_1^\perp(x,\Deltav) . \, 
\label{eq:spectrum_tmd}
\end{eqnarray}
%%%
where  we  have introduced the azimuthal angle $\phi$, which is defined via 
\begin{eqnarray}
\cos \phi = \frac{\Deltav \cdot \pv}{|\Deltav||\pv|} \, .
\end{eqnarray}
%%%%
{ Moreover,  $f_1$ is the unpolarized  TMD gluon distribution \footnote{Strictly speaking we should refer to it as the unintegrated gluon distribution, see e.g. the discussion in \cite{Altinoluk:2019fui}.} and  $h_1^\perp$ is the distribution of linearly polarized gluons in the nucleon.}
Clearly, this form of the cross section also applies in our case, we only have to keep in mind, that the relevant TMDs are defined with the staple-shaped Wilson line and that the { linearly polarized gluon distribution is identical to the unpolarized one.}
In our case we should substitute in Eq.(\ref{eq:spectrum_tmd})
%%%
\begin{eqnarray}
    xf_1(x,\Deltav) \to x G_{\rm dip}(x,\Deltav),\quad  xh^1_\perp(x,\Deltav) \to  x G_{\rm dip}(x,\Deltav).
\end{eqnarray}
%%%
Then, the hard matrix element, decomposed into an isotropic and an azimuthally modulated part reads:
%%%
\begin{eqnarray}
H_{qg \to G^* q} &=& \frac{ \alpha_s}{4 N_c} \, z^2  \overline{\sum_\text{pol.}} \partial^i_{p\perp} \Psi_{T,L}(z,\pv) \, \partial_{p \perp}^j \Psi_{T,L}^*(z,\pv) \, \delta_\perp^{ij} \, , \nonumber \\
\cos(2 \phi) \, H^{cos(2 \phi)}_{qg \to G^* q}
&=& \frac{ \alpha_s}{4 N_c} \, z^2  \overline{\sum_\text{pol.}}\partial^i_{p\perp} \Psi_{T,L}(z,\pv) \, \partial_{p \perp}^j \Psi_{T,L}^*(z,\pv) \, \Big( 2 \frac{\Delta^i \Delta^j}{\Deltav^2} - \delta_\perp^{ij} \Big) \, . 
%    M^j_{T,L} = \int d^2 \rr \, \exp(-i \pv \cdot \rr) \, r_j \Psi_{T,L}(z,\rr) = i \partial_{p_\perp}^j \Psi_{T,L}(z, \pv) .
\end{eqnarray}
%%%
To our knowledge, the simple relation between on-shell hard matrix elements and the derivatives of light front wave functions has not been stressed previously in the literature.

For completeness,  we present  in the appendix the  expressions for TMD factorization derived in the two - gluon exchange approximation, which applies for the case of hard gluons outside of a possible saturation regime.

Finally, let us show how to obtain the results in the back--to--back correlation limit from the formulas presented in Sections \ref{sec:formalism_dijet} and \ref{sec:particular}.
Namely, the results presented there indicate that the dependence of the spectrum on $ \pv$ and $\Deltav$ is determined by the functions
${\cal E}_{1,2}(\pv, z\Deltav,\epsilon)$. Expanding these functions in the limit of $\Deltav^2 \ll \pv^2$, we obtain:
%%%%
\begin{eqnarray}
    {\cal E}_1(\pv, z\Deltav,\epsilon) &=& \frac{1}{2} z^2 \,  \frac{4 p_i p_j}{[p^2 +\epsilon^2]^2} \Delta_i \Delta_j + \dots \, , \nonumber \\
     {\cal E}_2(\pv, z\Deltav,\epsilon) &=& \frac{1}{2} z^2 \,  \frac{1}{p^2 + \epsilon^2} \, \Big\{ \delta_{ij} - \frac{4 \epsilon^2}{[p^2 + \epsilon^2]^2} p_i p_j \Big\} \, \Delta_i \Delta_j + \dots \, .
\end{eqnarray}
%%%%
Using the relation of $f(x,\Deltav)$ to the dipole gluon TMD, we see that there emerges effectively a polarization density matrix of gluons
%%%%
\begin{eqnarray}
    f_{ij}(x,\Deltav) = {\Delta_i \Delta_j} \, f(x,\Deltav) = \frac{4 \pi^2 \alpha_s}{N_c} \, \, \frac{\Delta_i \Delta_j}{\Deltav^2} \, x  G_{\rm dip}(x,\Deltav)
\end{eqnarray}
%%%%
meaning that small-$x$ gluons are 100\% linearly polarized along their transverse momentum. 
This is of course in full agreement with Eq.(\ref{eq:F_and_H}). Furthermore, by inspecting the results of Subsection \ref{subsec:gamma_jet}, we immediately see that for the case of prompt photons and massless quarks, the $\cos(2\phi)$ modulation vanishes as indeed shown in Ref.\cite{Metz:2011wb}.
%For completeness, we give the relation of $f(x,\Deltav)$ to the dipole gluon TMD, which reads:
%%%
%\begin{eqnarray}
%f(x,\Deltav) = \frac{4 \pi^2 \alpha_s}{N_c} \, \frac{1}{\Deltav^2} \, x\frac{d G_{\rm dip}(x,\Deltav)}{d^2 \Deltav} \, .  \end{eqnarray}
%%%
We also wish to point out in closing, that the proportionality of the cross section to the dipole gluon TMD goes beyond the back-to-back correlation limit, as evidenced by the results in Sections \ref{sec:formalism_dijet} and \ref{sec:particular}. 
\

\section{Summary}
\label{sec:conc}
One of the main challenges of hadronic physics is the theoretical understanding of the QCD dynamics at high energies. The onset of nonlinear effects in the hadron wave function is expected at small values of the Bjorken $x$ variable, with the breakdown of the standard collinear factorization and the modification of the cross-section for the particle production at forward rapidities in hadronic collisions. In particular,   two - particle production is predicted to be sensitive to the description of the dense target. In the general case, when two partons are produced, the presence of nonlinear effects implies a complex structure for the cross-section, that involves quantities that are not probed in the single particle production. Moreover, the predictions for the two hadron production are affected by the hadronization. Such aspects  motivate the analysis of the associated jet plus  electroweak gauge boson production, since in this case the color neutral gauge boson does not suffer strong interactions and the cross-section is dependent on the same quantities present in the single electroweak boson production. In this paper, we have extended the formalism developed in Ref.  \cite{Bandeira:2024zjl}) for the associated jet + $G$ production and derived the general formula for the differential cross-section at the partonic level, which is the main input to estimate the corresponding observable. In particular, we have presented, for the first time, the expressions associated with the jet + $W^{\pm}$ case and demonstrated that the general formula reproduces the approximated expressions previously used in the literature to estimate the jet + $\gamma$ and jet + $Z^0$ production. { Moreover, we have established the relation of our results with the TMD factorization approach in the back - to - back correlation limit.}  The results presented in this paper strongly motivate the calculation of the differential cross-sections at the hadronic level and the comparison of the predictions with the current experimental data for the production of electroweak gauge bosons at forward rapidities. Such analysis is ongoing, and the results will be presented in a forthcoming publication.

%*****************************
\section*{Acknowledgments}
%*****************************
 V. P. G. would like to thank the members of the Institute of Nuclear Physics Polish Academy
of Sciences for their warm hospitality during the completion of this study. 
 Y.B.B.  and V.P.G. were  partially supported by CNPq, CAPES (Finance code 001), FAPERGS and  INCT-FNA (Process No. 464898/2014-5). The work of W.S. was partially supported by 
the Polish National
Science Center Grant No. UMO-2023/49/B/ST2/03665.

\appendix*
\section{Back-to-back correlation limit in the two-gluon exchange aproximation}

Here we demonstrate  how the hard scattering limit formulated in terms of the standard transverse--momentum  dependent (TMD) gluon distribution emerges directly from the presented formalism in the two-gluon exchange approximation.  
This is clearly only appropriate in a situation where one can 
neglect saturation corrections. The precise Wilson line structure is not of importance for the case discussed below.

Now, we should first expand the (anti-)quark-target $S$-matrices to the lowest order in the QCD eikonal. The $S$ - matrix of the $q\bar{q}$ - nucleon interaction, present in the definition of the dipole - nucleon cross -section [Eq. (\ref{Eq:dip})],  is defined by \cite{Nikolaev:2003zf,Nikolaev:2004cu,Nikolaev:2005dd,Nikolaev:2005zj,Nikolaev:2005ay,Nikolaev:2005qs}:
\begin{eqnarray}
    S_{q\bar{q}}(\bv_q,\bv_{\bar{q}}) = \frac{\langle N|\mbox{Tr}[S_{q}(\bv_q)S^{\dagger}_{q}(\bv_{\bar{q}})]|N\rangle}{\langle N|\mbox{Tr}[\openone_{N_c}]|N\rangle} \,\,,
\end{eqnarray}
where the trace is in the color space. At the leading order, the amplitude for the inelastic interaction is driven by the one - gluon exchange, which implies that in order to estimate the total cross - section for the dipole - nucleon interaction we have to consider the contribution of a color singlet two - gluon exchange in the $t$ - channel. In this approximation, the $S$ - matrices for the $qN$ and $\bar{q}N$ interactions will be given by
\begin{eqnarray}
  S_{q}(\bv_q) = \openone + iT^a \chi^a(\bv_q)  - \frac{1}{2}(T^a \chi^a (\bv_q) )^2 \,\,, \\
  S_{\bar{q}}(\bv_{\bar{q}}) = S^\dagger_{q}(\bv_{\bar{q}})= \openone - iT^a \chi^a (\bv_{\bar{q}})  - \frac{1}{2} (T^a\chi^a(\bv_{\bar{q}}))^2 \,\,, 
\end{eqnarray}
where $T^a$, $a = 1,\dots, N_c^2-1$, are the generators of $SU(N_c)$ in the fundamental representation. %$T_{\bar{q}}^\alpha \hat{V}_\alpha\chi(\bv)$ is the eikonal operator for the (anti-)quark-nucleon single - gluon interaction. 
It is important to emphasize that the above expansion satisfies the unitarity condition $S_{i}(\bv_i)S^{\dagger}_{i}(\bv_i) = I$, with $i = q, \bar{q}$, and that the eikonal function can be expressed in terms of the gluonic field as follows
%%%%
\begin{eqnarray}
    \chi^a(\bv) = g_s \int d x^+ \, A^{a -}(x^+,x^-,\bv)\Big|_{x^- = 0} \, ,
\end{eqnarray}
%%%%
where $g_s = \sqrt{4 \pi \alpha_s}$.
As a consequence, in this approximation, the dipole cross-section will be given by
%%%%
\begin{eqnarray}
\sigma_{q \bar q}(\rr) = C_F \int d^2\brho 
\frac{\Big \langle N \Big| \Big( \chi^a(\brho) - \chi^a(\brho + \rr) \Big)^2 \Big| N \Big\rangle}{\Big \langle N \Big| \mbox{Tr} \openone_{N_c^2-1} \Big| N\Big \rangle} =  2 C_F \int d^2\brho 
\frac{\Big \langle N \Big| \Big( \chi^a(\brho)\chi^a(\brho) - \chi^a(\brho)\chi^a(\brho + \rr) \Big) \Big | N \Big \rangle}{
\Big \langle N \Big| \mbox{Tr} \openone_{N_c^2-1} \Big| N\Big \rangle} \,, 
\label{eq:dip_tmdlimit}
\end{eqnarray}
where a sum over adjoint color indices $a$ is implied, and $C_F = (N_c^2 -1)/2N_c$ is the quadratic Casimir for the fundamental representation of the $SU(N_c)$ color group.

On the other hand, the TMD gluon distributions are defined through the matrix elements
%%%%%
\begin{eqnarray}
    &&\frac{1}{4 \pi^3 } \int d^2 \rr  d^2 \brho \,  \exp(-i \kv \cdot \rr) \, \frac{\Big \langle N \Big| \partial^i_\perp \chi^a(\brho) \, \partial^j_\perp \chi^a(\brho + \rr) \Big| N \Big\rangle}
    {\Big \langle N \Big| \mbox{Tr} \openone_{N_c^2-1} \Big| N\Big \rangle} = \frac{1}{2} \delta^{ij}_\perp \, xf_1(x,\kv)
    %\nonumber \\
    %&+& 
    + \frac{1}{2} \Big( 2 \frac{k^i k^j}{\kv^2} - \delta^{ij}_\perp \Big) xh_1^\perp(x,\kv) \, ,
    \label{eq:two-gluon}
\end{eqnarray}
%%%%%%
{ where  the unpolarized  TMD gluon distribution  $f_1$ is directly related to the corresponding collinear distribution as follows}
\begin{eqnarray}
g(x,\mu^2) = \int d^2\kv \, f_1(x,\kv,\mu^2) \, .     
\end{eqnarray}
%Moreover, $h_1^\perp$ is the distribution of linearly polarized gluons in the nucleon.
Using Eq. (\ref{eq:dip_tmdlimit}), we can establish the following relation between the dipole - nucleon cross - section and the TMD gluon distributions
%%%%%
%\begin{eqnarray}
%\sigma_{q \bar q}(\rr) = C_F \int d^2\brho 
%\frac{\Big \langle N \Big| \Big( \chi^a(\brho) - \chi^a(\brho + \rr) \Big)^2 \Big| N \Big\rangle}{\Big \langle N \Big| N\Big \rangle} =  2 C_F \int d^2\brho 
%\frac{\Big \langle N \Big| \Big( \chi^a(\brho)\chi^a(\brho) - \chi^a(\brho)\chi^a(\brho + \rr) \Big) \Big| N \Big\rangle}{\Big \langle N \Big| N\Big \rangle} \,. 
%\end{eqnarray}
%%%%
%we easily obtain a transverse momentum dependent density matrix of gluons as
%%%%
\begin{eqnarray}
\frac{N_c}{16 \pi^4 \alpha_s} \int d^2\rr \, \exp(-i \kv \cdot \rr) \, \partial^i_\perp \partial^j_\perp  \, \sigma_{q \bar q}(\rr) =  \frac{1}{2} \delta^{ij}_\perp \, xf_1(x,\kv) + \frac{1}{2} \Big( 2 \frac{k^i k^j}{\kv^2} - \delta^{ij}_\perp \Big) xh_1^\perp(x,\kv) \, . 
\end{eqnarray}
%%%%
Here we should notice, that just as in the main text, we have that  $f_1 = h_1^\perp$.
Let's now focus on the spectrum for the associated jet + electroweak gauge boson production. In the back - to - back correlation limit, where 
 $ |\Deltav| \ll |\pv|$, the hard process is concentrated in a region $ |\rr| \ll 1/|\pv|$ and the impact parameter variable $\sv$ is the difference of the location of the hard process in the impact parameter plane in amplitude and complex conjugate amplitude.
As a consequence, in order to derive the corresponding limit of eq. (\ref{eq:Master_formula_reduced}) one would expand the curly bracket in $\rr, \rr'$. To the second order in dipole sizes, we obtain:
%%%%%
\begin{eqnarray}
 \sigma_{q\bar{q}}(\sv+z\rr') + \sigma_{q\bar{q}}(\sv-z\rr) 
- \sigma_{q\bar{q}}(\sv -z(\rr-\rr') ) - \sigma_{q\bar{q}}(\sv) = \frac{z^2}{2} \Big( r_i r'_j + r_j r'_i \Big) \partial^i_\perp \partial^j_\perp \, \sigma_{q\bar{q}}(\sv) \,\,.
\end{eqnarray}
%%%%%
As a consequence,  our master formula reduces to
%%%%%
\begin{eqnarray}
   \frac{d\sigma^f_{T,L} (q_f \rightarrow G(p_G) q_k(p_q))}{dz d^2\pv d^2\Deltav} & = & \frac{1}{2 (2\pi)^4}\, 
\int  d^2\rr d^2\rr'  \exp[- i \pv \cdot (\rr - \rr^{\prime}) ] \Psi_{T,L}(z,\rr) \Psi^{*}_{T,L}(z,\rr') \frac{1}{2}\Big( r_i r'_j + r_j r'_i \Big)  \nonumber \\
&\times&\int d^2\sv \exp[-i \Deltav \cdot \sv] 
%\nonumber \\ 
\partial^i_\perp \partial^j_\perp \, \sigma_{q\bar{q}}(\sv) 
\nonumber \\
&=& \frac{1}{2 (2 \pi)^4} \, \frac{z^2}{2}
\Big( \partial^i_{p\perp} \Psi(z,\pv) \, \partial_{p \perp}^j \Psi^*(z,\pv) + (i \leftrightarrow j) \Big) \, \int d^2\sv \exp[-i \Deltav \cdot \sv] 
%\nonumber \\ 
\partial^i_\perp \partial^j_\perp \, \sigma_{q\bar{q}}(\sv)
\nonumber \\
&=& H_{qg \to G^*q}\, x f_1(x,\Deltav) + \cos(2 \phi) \, H^{cos(2 \phi)}_{qg \to G^* q}
\, x h_1^\perp(x,\Deltav) \, .
\label{eq:spectrum_tmd2}
%&=& \frac{\alpha_s}{4 \pi^2 \, N_c} \Big\{ |\bM_{T,L}|^2 \, x f_1(x,\Deltav) + \frac{1}{2} \Big( \frac{|\bM_{T,L} \cdot \Deltav|^2}{\Deltav^2} - |\bM_{T,L}|^2 \Big) \, xh^\perp_1(x,\Deltav) \Big\} \, , 
\end{eqnarray}
%%%%
Just as in the case described in the main text, also here there is in fact only one independent unintegrated gluon distribution.
The reason for this is that the gluon field strength tensor is dominated by $\partial^i_\perp A^{-a}$, which in the approximation of Eq.(\ref{eq:two-gluon}) translates to the effective linear gluon polarization boson along its transverse momentum.

%%%%%%%%%%%%%%%%%%%%%
\bibliographystyle{unsrt}

\end{document}